\documentclass[nofootinbib]{revtex4}
\usepackage{graphicx}
\usepackage{latexsym}
\def\be{\begin{equation}}
\def\ee{\end{equation}}
\def\bea{\begin{eqnarray}}
\def\eea{\end{eqnarray}}

\begin{document}

\hfill USTC-ICTS-14-19

\title{Entropic mechanisms with generalized scalar fields in the Ekpyrotic universe}

\author{Mingzhe Li}
\email{limz@ustc.edu.cn}
\affiliation{Interdisciplinary Center for Theoretical Study, University of Science and Technology of China, Hefei, Anhui 230026, China}
\affiliation{State Key Laboratory of Theoretical Physics, Institute of Theoretical Physics, Chinese Academy of Sciences, Beijing 100190, China}

\begin{abstract}

For the Ekpyrotic universe, the entropic mechanisms with minimal couplings, which have been used to generate nearly scale invariant primordial
perturbations, was proved to be unstable. To overcome this difficulty, some non-minimal coupling entropic models were proposed. In this paper
we extend these studies to the cases where all the scalar fields have non-standard kinetic terms. We first prove that in these general cases,
without non-minimal couplings the entropic models are still unstable. The condition for the stability conflicts with the requirement for
achieving scale invariant perturbations. Then we study the non-minimal coupling models and show that at least for some simple cases these
models are stable and able to generate the primordial perturbations consistent with current observations.

\end{abstract}

\maketitle

\hskip 1.6cm PACS number(s): 98.80.Cq, 98.80.Bp \vskip 0.4cm

\section{Introduction}

Inflationary universe \cite{Inflation} provides not only solutions to the horizon and flatness problems of the hot big bang cosmology but also
mechanisms to produce the primordial perturbations which seed the large scale structure. Recent observations from the Planck satellite
\cite{Planck1,Planck} confirmed that the primordial density perturbation is adiabatic, nearly scale-invariant and satisfies Gaussian statistics.
These results are highly consistent with the predictions of the simple inflation models.

Even though the success of inflation, there exist alternative pictures in the literature. One of which is the Ekpyrotic/Cyclic model
\cite{Ekpyrotic}. In the Ekpyrotic model the universe is assumed to have experienced a slow contracting phase (Ekpyrotic phase) before bouncing
to the hot expansion. A slow contraction driven by stiff matter (the equation of state $w>1$) is needed to explain the smoothness and flatness
of the universe and to suppress the BKL anisotropies \cite{BKL}. Such stiff matter can be modeled by a scalar field with a canonical kinetic
term and a negative but steep potential, for instance the frequently studied model in which the scalar field has a negative exponential
potential and is minimally coupled to the Einstein's gravity
\be\label{single}
S=\int d^4x \sqrt{g} [\frac{R}{2}+\frac{1}{2}\partial_{\mu}\phi\partial^{\mu}\phi+V_0 \exp(-c\phi)]~,
\ee
where $V_0$ and $c$ are positive constants, and we have used the most negative signature for the metric and the unit $8\pi G=1$.
This model is invariant under the scale transformation $g_{\mu\nu}\rightarrow \omega^2 g_{\mu\nu},~\phi\rightarrow \phi+2(\ln\omega)/c$,
here $\omega$ is independent of the spacetime. So it has a scaling solution in which the equation of state is a constant, $w=c^2/3-1$,
which is larger than one if $c>\sqrt{6}$. The productions of super-Hubble density perturbations are due to the fact that during the contraction
the Hubble radius was shrinking and the quantum vacuum fluctuations created deep inside it were able to cross the Hubble radius to outside
regions. However, it was shown \cite{Ekpyrotic4} that the single field Ekpyrotic model (\ref{single}) cannot produce nearly scale-invariant
density perturbation, the spectrum of the curvature perturbation is strongly blue tilted and ruled out by the observations.

Currently the best way around this problem is the entropic mechanism in which multiple fields are introduced.
In the entropic mechanism the scalar perturbations during the Ekpyrotic phase are dominated by the entropy (or isocurvature) modes.
In fact the adiabatic mode during this time is not relevant because it has a strongly blue tilted spectrum and is suppressed deeply at large
scales. Some of the entropy perturbations may have scale invariant spectra. The adiabatic perturbation seeds the large scale structure is
regarded as being converted from the scale invariant entropy perturbation at some time later than the Ekpyrotic phase. Such a conversion will
not change the shape of the spectrum. A simple entropic model can be constructed by multi canonical and minimally coupled scalar fields,
for example the model of double fields with similar negative exponential potentials
\cite{Ekpyroticentropy},
\be\label{double}
S=\int d^4x \sqrt{g} [\frac{R}{2}+\frac{1}{2}\partial_{\mu}\phi_1\partial^{\mu}\phi_1+
\frac{1}{2}\partial_{\mu}\phi_2\partial^{\mu}\phi_2+V_1 \exp(-c_1\phi_1)+V_2\exp(-c_2\phi_2)]~.
\ee
This model also has the scale symmetry
 $g_{\mu\nu}\rightarrow \omega^2 g_{\mu\nu},~\phi_1\rightarrow \phi_1+2(\ln\omega)/c_1,~\phi_2\rightarrow \phi_2+2(\ln\omega)/c_2$ and
admits the scaling solution with $w=c^2/3-1,~c^2=c_1^2c_2^2/(c_1^2+c_2^2)$, which is important for the slow contraction.
It was shown \cite{Ekpyroticentropy} that with this scaling solution, scale-invariant entropy perturbation can be achieved in the limit
$w\gg 1$, this requires both $c_1$ and $c_2$ are large enough, i.e., both scalar fields have steep potentials. However this requirement
brings new difficulties. First it was pointed out in Refs. \cite{koyama} that the scaling solution is unstable, the entropy direction has
a tachyonic effective potential. Second
the steep potentials results in large non-Gaussianity during the Ekpyrotic phase
\cite{Koyama:2007if,Buchbinder:2007at,Lehners:2007wc,Lehners:2008my} and may conflict with the observations.
In all, when confronting these simple entropic models with the observational data, we need to fine tune the model parameters and the initial
conditions at the beginning of the Ekpyrotic phase.

To overcome these difficulties, a new entropic mechanism \cite{Li:2013hga} was proposed in which the second field is massless but its kinetic
term has a non-minimal coupling to the first field, for example
\be\label{action2}
S=\int d^4x \sqrt{g}[\frac{R}{2}+\frac{1}{2}\partial_{\mu}\phi_1\partial^{\mu}\phi_1+V_1 \exp(-\lambda\phi_1)+
{1\over 2}\exp(-\alpha\phi_1)\partial_{\mu}\phi_2\partial^{\mu}\phi_2]~.
\ee
In this mechanism the first field $\phi_1$ dominates the universe and drives the Ekpyrotic contraction. The second field $\phi_2$ serves as
a spectator, it is frozen during the Ekpyrotic phase due to the friction brought by the non-minimal coupling and always represents the
entropy direction. But its fluctuation (the entropy perturbation) gets amplified. Given above exponential potential and exponential coupling,
the Ekpyrotic phase is realized by the scaling solution with $w=\lambda^2/3-1$, the spectator behaves as a massless field living in an
effective de Sitter space if $\lambda=\alpha$, so that its vacuum fluctuation generates the scale-invariant entropy perturbation. Notice
that when $\lambda=\alpha$, the above action (\ref{action2}) is invariant under the scale transformation
$g_{\mu\nu}\rightarrow \omega^2 g_{\mu\nu},~\phi_1\rightarrow \phi_1+2(\ln\omega)/\lambda,~\phi_2\rightarrow \omega\phi_2+C$ for arbitrary constant
$C$. The adiabatic perturbation generated by $\phi_1$ has a blue spectrum and can be neglected at large scales.
After the bounce, the scale-invariant entropy perturbation can be converted into the scale-invariant adiabatic perturbation by some
mechanisms such as the curvaton \cite{curvaton} or modulated preheating \cite{modulation1,modulation2,modulation3}.
This is similar to the Conformal \cite{conformal1} universe, pseudo Conformal universe \cite{conformal2} and the Galileon
Genesis \cite{galileon} (see also the slow expansion scenario for different case \cite{piao}).
Such non-minimal coupling mechanism has some advantages compared with the old one. First, the scaling solution is stable and we do not need to
fine tune the initial condition of the Ekpyrotic phase. Second, it was proved in Ref. \cite{Fertig:2013kwa} that the non-Gaussianity is small
and consistent with current observations. Third, the requirement $w\gg 1$ in the old entropic mechanism can be relaxed, we only need $w>1$ to
suppress the anisotropies, this liberates the model buildings. In Ref. \cite{Ijjas:2014fja}, the mechanism was studied in more general case
where the exponential potential and exponential coupling function are replaced by more general functions, i.e.,
\be
S=\int d^4x \sqrt{g}[\frac{R}{2}+\frac{1}{2}\partial_{\mu}\phi_1\partial^{\mu}\phi_1+V(\phi_1)+
{1\over 2}\Omega^2(\phi_1)\partial_{\mu}\phi_2\partial^{\mu}\phi_2]~.
\ee
Note that similar mechanism was considered within the curvaton scenario \cite{qiu1} and applied to the non-singular bouncing universe
\cite{qiu2}. Other discussions or applications of this non-minimal coupling entropic mechanism can be found in Refs.
\cite{Koehn:2013upa,Feng:2013pba,Feng:2014tka,Battarra:2014tga,Battefeld:2014uga,Battarra:2014xoa,Battarra:2014kga}.

Up to now all the studies focused on the Ekpyrotic models where all the scalar fields have the canonical kinetic terms. In this paper we will
study the entropic mechanism of the Ekpyrotic/Cyclic universe in more general cases where the kinetic terms of the multi scalar fields have
non-standard forms.
In cosmology the scalar fields with non-standard kinetic terms had been applied to the theories of inflation \cite{kinflation} and dark energy
\cite{kessence}. First we will show that even with  non-standard kinetic terms, the model with minimal coupling is still unstable, this means that the conclusion of \cite{koyama} is robust.
Then we will discuss the non-minimal coupling model with non-standard kinetic terms.

\section{Entropic mechanism with minimal couplings}

In this section we will generalize the analysis of \cite{koyama} to the cases where the scalar fields have non-standard
kinetic terms. For simplifying the analysis, the model what we consider contains two fields with minimal couplings
\be\label{action1}
S=\int d^4x\sqrt{g}[\frac{R}{2}+\frac{K_1(X_1)}{\phi_1^2}+\frac{K_2(X_2)}{\phi_2^2}]~,
\ee
where $X_{\alpha}=1/2\partial_{\mu}\phi_{\alpha}\partial^{\mu}\phi_{\alpha},~\alpha=1,2$ and $K_{\alpha}$ only depends on $X_{\alpha}$.
It is invariant under the scale transformation $g_{\mu\nu}\rightarrow \omega^2 g_{\mu\nu},~\phi_{\alpha}\rightarrow \omega\phi_{\alpha}$.
The scalar field with canonical kinetic term and exponential potential can be considered as the case where the function $K_{\alpha}$ has a
special form, i.e., for the scalar field with
\be
\mathcal{L}=X_{\psi}+V_0\exp(-c\psi)~,
\ee
this Lagrangian density can be rewritten as
\be
\mathcal{L}=\frac{(4/c^2)(X_{\phi}+V_0)}{\phi^2}
\ee
through the field redefinition $\psi=(2/c)\ln (c\phi/2)$.

In the the spatially flat Friedmann-Robertson-Walker (FRW) universe,
\be\label{FRW}
ds^2=a^2(\eta) \eta_{\mu\nu} dx^{\mu}dx^{\nu}~,
\ee
it is not difficult to get the equations of motion including the Friedmann equation from the action (\ref{action1})
\bea\label{eom}
& &\phi_{\alpha}'' \frac{\partial\rho_{\alpha}}{\partial X_{\alpha}}+2\mathcal{H}\phi_{\alpha}'(\frac{\partial P_{\alpha}}{\partial X_{\alpha}}
-X_{\alpha}\frac{\partial^2 P_{\alpha}}{\partial X_{\alpha}^2})=\frac{2a^2}{\phi_{\alpha}}\rho_{\alpha}\nonumber\\
& &\mathcal{H}^2=\frac{a^2}{3}(\rho_1+\rho_2)~,
\eea
where the pressure is $P_{\alpha}=K_{\alpha}/\phi_{\alpha}^2$, the energy density is
$\rho_{\alpha}=2X_{\alpha} (\partial P_{\alpha}/\partial X_{\alpha})-P_{\alpha}$,
and the repeated subscripts do not mean the summation here. The primes denote the derivatives with respect to the conformal time $\eta$
and $\mathcal{H}=a'/a$ is the reduced Hubble parameter. Following Ref. \cite{kessence}, we define the variables
\be
x_{\alpha}=\sqrt{\Omega_{\alpha}}\geq 0,~y_{\alpha}=\sqrt{X_{\alpha}}\geq 0~,
\ee
where $\Omega_{\alpha}\equiv \rho_{\alpha}/(\rho_1+\rho_2)$ is the density parameter, and for the two species universe $x_1$ and $x_2$ are not independent,
\be
x_1^2+x_2^2=1~.
\ee
We can choose the three independent variables $x_1$, $y_1$ and $y_2$ and rewrite the equations of motion (\ref{eom}) as
\bea\label{eom1}
\frac{dx_1}{dN}&=&\frac{3}{2}x_1(1-x_1^2)(w_2-w_1)\nonumber\\
\frac{dy_{\alpha}}{dN}&=&\frac{\sigma_{\alpha} 2\sqrt{6}\sqrt{\tilde{\rho}_{\alpha}}x_{\alpha}-3\dot K_{\alpha}}{\ddot K_{\alpha}}~,
\eea
where $N=\ln a$ is the e-folding number, $\dot K_{\alpha}=dK_{\alpha}/dy_{\alpha}$, $\tilde{\rho}_{\alpha}=-K_{\alpha}+y_{\alpha}\dot K_{\alpha}
=\rho_{\alpha}\phi_{\alpha}^2$. The notation
$\sigma_{\alpha}=\pm 1$ represents the sign of the combination $\mathcal{H}\phi'_{\alpha}/\phi_{\alpha}$, we will consider the plus sign for
both fields for simplicity. This means  $\phi'_{\alpha}/\phi_{\alpha}<0$ because $\mathcal{H}<0$ during the Ekpyrotic phase.
The equations of state appeared in above equations are expressed as
\be
w_{\alpha}=\frac{K_{\alpha}}{\tilde{\rho}_{\alpha}}~,
\ee
and only depend on $y_{\alpha}$.
Furthermore, we have the sound speed for each field
\be
c_{s\alpha}^2=\frac{\dot K_{\alpha}}{\dot{\tilde{\rho}}_{\alpha}}=\frac{\dot K_{\alpha}}{y_{\alpha}\ddot K_{\alpha}}~,
\ee
which will be useful in the following discussions.

\subsection{Scaling solution}

The equations (\ref{eom1}) can be further rewritten as
\bea\label{equations}
\frac{dx_1}{dN}&=&\frac{3}{2}x_1(1-x_1^2)(w_2-w_1)\nonumber\\
 \frac{dy_1}{dN}&=&{3\over 2}\frac{1-w_1}{\dot r_1}(x_1-r_1)\nonumber\\
  \frac{dy_2}{dN}&=&{3\over 2}\frac{1-w_2}{\dot r_2}(x_2-r_2)~,
 \eea
here following Ref. \cite{kessence} we defined the function $r_{\alpha}$ as
\be
r_{\alpha}(y_{\alpha})=\frac{3\dot K_{\alpha}}{2\sqrt{6}\sqrt{\tilde{\rho}_{\alpha}}}~,
\ee
so that its derivative can be expressed as
\be
\dot r_{\alpha}=\frac{dr_{\alpha}}{dy_{\alpha}}=\frac{3\ddot{K}_{\alpha}}{4\sqrt{6}\sqrt{\tilde{\rho}_{\alpha}}}(1-w_{\alpha})~.
\ee

There are several critical points obtained from the equations (\ref{equations}) for which $dx_1/dN=0$, and $dy_{\alpha}/dN=0$.
But only the following one is of interest for the Ekpyrotic phase,
\be\label{critical}
w_1(y_{10})=w_2(y_{20})=w_0, x_{10}=r_1(y_{10})\equiv r_{10}, x_{20}=r_2(y_{20})\equiv r_{20}~,
\ee
where $w_0$, $r_{10}$ and $r_{20}$ are constants and of course we have $r_{10}^2+r_{20}^2=1$. This critical point corresponds to the solution
where both fields have the same equation of state and their energy densities scales with the same rate. Ekpyrotic phase requires $w_0>1$.
The next step is to investigate whether this solution is stable. To do that we should make small perturbations around this solution and solve
the linear equations for the perturbations. In terms of the relation
\be
\frac{dw_{\alpha}}{dy_{\alpha}}=
\frac{2(1+w_{\alpha})}{1-w_{\alpha}}\frac{\dot r_{\alpha}}{r_{\alpha}}(c_{s\alpha}^2-w_{\alpha})~,
\ee
we obtain the following linear equations around the scaling solution (\ref{critical}),
\be\label{eom2}
\frac{d}{dN}\left(
\begin{array}{c}
\delta x_1\\
\delta y_1\\
\delta y_2
\end{array}
\right)=\left(
\begin{array}{ccc}
0 & B & C\\
D & E & 0\\
F & 0 & E
\end{array}\right)\left(\begin{array}{c}
\delta x_1\\
\delta y_1\\
\delta y_2
\end{array}
\right)
\ee
where the elements of the coefficient matrix are listed below
\bea\label{coeff}
& &B=-\frac{3(1+w_0)\dot r_{10}}{1-w_0}r_{20}^2(c_{s1}^2-w_0)~,
~C=\frac{3(1+w_0)\dot r_{20}}{1-w_0}r_{10}r_{20}(c_{s2}^2-w_0)~,\nonumber\\
& &D=\frac{3}{2}\frac{1-w_0}{\dot r_{10}}~,~E=\frac{3}{2}(w_0-1)~,~F=-\frac{3}{2}\frac{r_{10}}{r_{20}}
\frac{1-w_0}{\dot r_{20}}~.
\eea
The eigenvalues of the coefficient matrix can be obtained by solving the following equation
\be
\left|
\begin{array}{ccc}
-e & B & C\\
D & E-e & 0\\
F & 0 & E-e
\end{array}\right|=0~.
\ee
We find that this equation has a very simple form
\be
(e-E)\left[e^2-Ee-(BD+CF)\right]=0~.
\ee
It has three roots
\be
e=E,~\frac{E\pm \sqrt{E^2+4(BD+CF)}}{2}~.
\ee
In the contracting universe, $N$ decreases, stability of the scaling solution requires the real parts of all the eigenvalues are positive.
So this requires
\be
E>0~{\rm and}~BD+CF<0~.
\ee
In terms of the expressions (\ref{coeff}), one obtains the stability condition
\be\label{stability}
1<w_0<r_{20}^2c_{s1}^2+r_{10}^2c_{s2}^2~,
\ee
where we have used $r_{10}^2+r_{20}^2=1$. This means that given a specific model (\ref{action1}), only if the above requirements are satisfied
the solution (\ref{critical}) is an attractor to support the Ekpyrotic phase.

\subsection{Entropy perturbation}

Now we consider the primordial perturbations generated in this Ekpyrotic model. For this double field model, the perturbations are decomposed
into the adiabatic and entropy modes. During the Ekpyrotic phase the adiabatic perturbation is suppressed on large scales, so we will focus
on the entropy perturbation. Because the whole system has no anisotropic stress at the linear level, it is convenience for us to use the
conformal Newtonian gauge
\be
ds^2=a^2[(1+2\Phi)d\eta^2-(1-2\Phi)\delta_{ij}dx^idx^j]~.
\ee
The linear equations from the energy and momentum conservation laws have the following forms \cite{Ma:1995ey,Li:2010hm}
\bea
\delta_{\alpha}'&=&(1+w_{\alpha})(3\Phi'-k^2v_{\alpha})+3\mathcal{H}(w_{\alpha}-c_{s\alpha}^2)\delta_{\alpha}+9\mathcal{H}^2(1+w_{\alpha})
(c_{a\alpha}^2-c_{s\alpha}^2)v_{\alpha}~,\nonumber\\
v_{\alpha}'&=&\mathcal{H}(3c_{s\alpha}^2-1)v_{\alpha}+\frac{c_{s\alpha}^2}{1+w_{\alpha}}\delta_{\alpha}+\Phi~,
\eea
where $v_{\alpha}=\delta\phi_{\alpha}/\phi_{\alpha}'$ and another sound speed (dubbed adiabatic sound speed in the literature)
$c_{a\alpha}^2=p_{\alpha}'/\rho_{\alpha}'=w_{\alpha}-w_{\alpha}'/[3\mathcal{H}(1+w_{\alpha})]$ is introduced, which is different from the true
sound speed $c^2_{s\alpha}$ for the scalar fields. With these two equations, we may get a second order equation
\bea\label{simplify}
v_{\alpha}''&=&[\mathcal{H}(3c_{a\alpha}^2-1)+2\frac{c_{s\alpha}'}{c_{s\alpha}}]v_{\alpha}'+[-c_{s\alpha}^2k^2
+2\mathcal{H}\frac{c_{s\alpha}'}{c_{s\alpha}}(1-3c_{s\alpha}^2)+3c_{s\alpha}^2(\mathcal{H}'-\mathcal{H}^2)
+3\mathcal{H}^2c_{a\alpha}^2-\mathcal{H}'+6\mathcal{H}c_{s\alpha}c_{s\alpha}']v_{\alpha}\nonumber\\
&+&3c_{s\alpha}^2(\Phi'+\mathcal{H}\Phi)+\Phi'-(2\frac{c_{s\alpha}'}{c_{s\alpha}}+3\mathcal{H}c_{a\alpha}^2)\Phi~.
\eea
Consider the scaling solution (\ref{critical}) in which $c_{s\alpha}^2$ are constants and $w_1=w_2=w_0=const.$, and the
adiabatic sound speed $c_{a\alpha}^2=w_{\alpha}=w_0$, the above equation has a rather simple form
\be\label{simpleform}
v_{\alpha}''=\mathcal{H}(3w_0-1)v_{\alpha}'+[-c_{s\alpha}^2k^2+3c_{s\alpha}^2(\mathcal{H}'-\mathcal{H}^2)
+3\mathcal{H}^2w_0-\mathcal{H}']v_{\alpha}
+3c_{s\alpha}^2(\Phi'+\mathcal{H}\Phi)+\Phi'-3\mathcal{H}w_0\Phi~.
\ee

The projections of the field perturbations $v_{\alpha}$ to the adiabatic direction and entropy direction have been done for multi canonical
scalar fields in Ref. \cite{Gordon:2000hv}. The generalizations to the non-standard scalar fields were discussed, for instances, in Ref.
\cite{Langlois:2008mn}. For the models considered in this paper we need only make a slight generalization of the method developed in
\cite{Gordon:2000hv}. The adiabatic direction $\sigma$ at the background is defined to be
\be
\frac{\sigma'^2}{a^2}=2X_1P_{X_1}+2X_2P_{X_2}=(\rho_1+P_1)+(\rho_2+P_2)=\rho+P~,
\ee
where $P_{X_{\alpha}}$ means the derivative of the total pressure with respect to $X_{\alpha}$.
So
\be
\sigma'^2=\phi_1'^2P_{X_1}+\phi_2'^2P_{X_2}~, {\rm and}~\sigma'=\cos\theta \sqrt{P_{X_1}}\phi'_1+\sin\theta\sqrt{P_{X_2}}\phi'_2~,
\ee
where the angle is
\be\label{angle}
\cos\theta=\frac{\phi_1'\sqrt{P_{X_1}}}{\sigma'}=\sqrt{\frac{\rho_1+P_1}{\rho+P}}~,~\sin\theta
=\frac{\phi_2'\sqrt{P_{X_2}}}{\sigma'}=\sqrt{\frac{\rho_2+P_2}{\rho+P}}~,
\ee
both fields are required to satisfy the null energy condition so that $P_{X_{\alpha}}\geq 0$.
The adiabatic perturbation represents the perturbation along the background trajectory
\be
\delta \sigma=\cos\theta \sqrt{P_{X_1}}\delta\phi_1+\sin\theta\sqrt{P_{X_2}}\delta\phi_2~,
\ee
hence
\be
\frac{\delta\sigma}{\sigma'}=\cos^2\theta v_1+\sin^2\theta v_2~.
\ee
The entropy direction is orthogonal to the adiabatic direction, so the entropy perturbation is
\be\label{entropyperturbation}
\delta s=\cos\theta \sqrt{P_{X_2}}\delta\phi_2-\sin\theta \sqrt{P_{X_1}}\delta\phi_1=\sigma'\cos\theta\sin\theta(v_2-v_1)~.
\ee
One can check that the entropy perturbation is automatically gauge-invariant.

In terms of the projections and the perturbed Einstein equation
\be
\Phi'+\mathcal{H}\Phi=-(\mathcal{H}'-\mathcal{H}^2)(\cos^2\theta v_1+\sin^2\theta v_2)~,
\ee
we can obtain the equation of motion for the entropy perturbation from Eq. (\ref{simpleform})
\be\label{entropyper}
\left.(\frac{\delta s}{\sigma'}\right.)''=\mathcal{H}(3w_0-1)\left.(\frac{\delta s}{\sigma'}\right.)'+
[3\mathcal{H}^2w_0-\mathcal{H}'+(3\mathcal{H}'-3\mathcal{H}^2-k^2)c_s^2]\left.(\frac{\delta s}{\sigma'}\right.)
+\sin\theta\cos\theta k^2(c_{s1}^2-c_{s2}^2)\left.(\frac{\delta \sigma}{\sigma'}\right.)~,
\ee
where we have defined the total sound speed
\be
c_s^2\equiv c_{s1}^2\sin^2\theta+c_{s2}^2\cos^2\theta~.
\ee
We see that even for constant $\theta$, the entropy perturbation is not decoupled from the adiabatic perturbation
due to the difference of the sound speeds of the two scalar fields.
With the scaling solution, we have the following expressions\footnote{We have assumed the conformal time $\eta$ is negative during the Ekpyrotic phase. }
\be\label{Hubble}
\mathcal{H}=\frac{2}{1+3w_0}\frac{1}{\eta}~,~\mathcal{H}'=-\frac{2}{1+3w_0}\frac{1}{\eta^2}~,
\ee
with these equations and through the definition
\be
u=a^{\frac{1-3w_0}{2}} \frac{\delta s}{\sigma'}~,
\ee
Eq. (\ref{entropyper}) can be rewritten as
\be\label{e2}
u''+[c_s^2k^2-\frac{2}{\eta^2}(1-\frac{9c_s^2(1+w_0)}{(1+3w_0)^2})]u=
a^{\frac{1-3w_0}{2}}\sin\theta\cos\theta k^2(c_{s1}^2-c_{s2}^2)\frac{\delta\sigma}{\sigma'}~.
\ee

As mentioned before, the adiabatic perturbation generated during the Ekpyrotic phase can be neglected. With this approximation, the equation
for the entropy perturbation becomes homogeneous, and the adiabatic perturbation required by the structure formation is converted from the
entropy perturbation. From the left hand side of Eq. (\ref{e2}) we know that the scale invariance of the power spectrum of the entropy
perturbation can be only achieved if $9c_s^2(1+w_0)/(1+3w_0)^2=0$. Slow roll inflation is one possibility because $1+w_0\simeq 0$.
Naively there is another possibility in which $w_0\rightarrow +\infty$, however if we take this limit we find that
\be
\frac{9c_s^2(1+w_0)}{(1+3w_0)^2}=9\frac{c_s^2}{w_0}\frac{1+\frac{1}{w_0}}{9+\frac{6}{w_0}+\frac{1}{w_0^2}}\rightarrow \frac{c_s^2}{w_0}~.
\ee
For the model we considered in this section, the total sound speed is
\be
c_s^2=c_{s1}^2\sin^2\theta+c_{s2}^2\cos^2\theta=c_{s1}^2\frac{\rho_2+P_2}{\rho+P}+c_{s2}^2\frac{\rho_1+P_1}{\rho+P}
=c_{s1}^2r_{20}^2+c_{s2}^2r_{10}^2~.
\ee
The stability condition (\ref{stability}) for the scaling solution obtained previously requires $c_s^2>w_0$ and conflicts with the
requirement of the scale-invariance, i.e., $c_s^2/w_0\rightarrow 0$. Hence in the Ekpyrotic universe, even considering non-standard kinetic
terms the minimal coupling entropic models are not stable when the requirement of scale invariant entropy perturbation is imposed.

\section{The entropic mechanism with non-minimal couplings}

Now in this section we will present some considerations on the entropic mechanism with non-minimal couplings. This is a generalization of the
model proposed in Ref. \cite{Li:2013hga} to the cases where the scalar fields have non-standard kinetic terms.
Roughly the idea is that a non-canonical field with the Lagrangian $\mathcal{L}=K_1(X_1)/\phi_1^2$ dominating the Ekpyrotic universe will
render the background evolution satisfying the scaling solution (\ref{Hubble}) with constant $w_1$ and $X_1$. Here the function $K_1(X_1)$
is assumed to contain a constant term so that it includes the normal Ekpyrotic model with a canonical kinetic term and a exponential potential as discussed
in the first paragraph of section II.
From the Friedmann equation one can obtain $\phi_1=-ha\eta$, where the constant $h>0$ if we assume $\phi_1>0$. Within this background,
a massless field $\phi_2$ non-minimally coupled to $\phi_1$ through the action
\be
S_{\phi_2}=\int d^4x \sqrt{g} \frac{1}{\phi_1^2} X_2
\ee
will feel that its fluctuations live in an effective de Sitter space, i.e., the quadratic action for the perturbation $\delta\phi_2$ is
\be\label{effectiveaction2}
S_{\delta\phi_2}={1\over 2}\int d^4x \frac{1}{h^2\eta^2}\eta^{\mu\nu}\partial_{\mu}\delta\phi_2\partial_{\nu}\delta\phi_2~,
\ee
where $\eta^{\mu\nu}$ is the Minkowski metric and $h$ can be considered as the effective Hubble constant. So that $\delta\phi_2$ will obtain
a scale invariant power spectrum $\mathcal{P}^{1/2}_{\delta\phi_2}= h/2\pi$ from the initial Bunch-Davies vacuum. This is very similar to
the Conformal and pseudo-Conformal universe \cite{conformal1,conformal2,galileon}.

More specifically we will study the following model
\be\label{nonminimal}
S=\int d^4x\sqrt{g}[\frac{R}{2}+\frac{K_1(X_1)}{\phi_1^2}+\frac{K_2(X_2)}{\phi_1^{\alpha}}]~,
\ee
where the spectator $\phi_2$ is shift symmetric but its kinetic term is non-minimally coupled to $\phi_1$. The parameter $\alpha$ is assumed
to be positive, if $\alpha=2$ the scale symmetry is recovered. More generally one can think that $K_2$ also depends on higher derivative
terms like $\Box\phi_2$. In this paper we will not consider this possibility and leave this to the future work.
With the same philosophy in Ref. \cite{Li:2013hga}, we require the second field has the frozen background $X_2=0$, and both its pressure
$P_2=K_2/\phi_1^{\alpha}$ and energy density $\rho_2=(2X_2/\phi_1^{\alpha})(dK_2/dX_2)-K_2/\phi_1^{\alpha}$ are sufficiently small within
this background. The first derivative at the frozen point $(dK_2/dX_2)|_{X_2=0}$ must be positive to avoid ghost instability, we will denote
it as $K_0$. The universe was contracting with the scaling solution
\be
X_1=const., ~w_1=const.,~\mathcal{H}=\frac{2}{1+3w_1}\frac{1}{\eta}~,
\ee
where $w_1$ only depends on $X_1$ and should be larger than one so that the contraction is slow.
With these considerations, the background equation of motion of $\phi_2$ is
\be
\phi_2''+(2\mathcal{H}-\alpha\frac{\phi_1'}{\phi_1})\phi_2'=0~.
\ee
As mentioned before, $\phi_1'/\phi_1<0$, in terms of the scaling solution and the Friedmann equation we obtain that the above equation is
\be
\phi_2''+(\alpha\sqrt{\frac{6X_1w_1}{K_1}}-2)(-\mathcal{H})\phi_2'=0~.
\ee
Because $\mathcal{H}<0$, one can see that the background of $\phi_2$ is indeed frozen at $\phi_2'=0$ as long as $\alpha\sqrt{6X_1w_1/K_1}>2$.
Since $\phi_2$ is a spectator, it has no contribution to the energy budget, whether the scaling solution is stable only depends on the model
of $\phi_1$. In terms of the similar method used in previous section, one can show that the scaling solution is indeed stable as long as
$w_1>1$ and $r_{1}=1$ can be satisfied. The perturbation of the spectator is always along the entropy direction. The projection angle
(\ref{angle}) of the perturbations $\theta=0$ and the entropy perturbation defined in (\ref{entropyperturbation}) is
$\delta s=\sqrt{dK_2/dX_2}\phi_1^{-\alpha/2}\delta \phi_2$.

It is straight forwardly to obtain the following quadratic action of the perturbation of the spectator $\delta\phi_2$ around the background,
\bea
S_{\delta\phi_2}=\frac{K_{0}}{2}\int d^4x \frac{a^2}{\phi_1^{\alpha}}[\delta\phi_2'^2-c_{s2}^2\partial_i\delta\phi_2\partial_i\delta\phi_2]
~,
\eea
where the sound speed is
\be
c_{s2}^2=\frac{K_0}{K_0+2X_2 (d^2K_2/dX_2^2)|_{X_2=0}}=1~.
\ee
In terms of the relations $\phi_1\propto -a\eta$ and $\mathcal{H}=2/[(1+3w_1)\eta]$, the quadratic action can be rewritten as
\be
S_{\delta\phi_2}
={1\over 2}\int d^4x q^2 \eta^{\mu\nu}\partial_{\mu}\delta\phi_2\partial_{\nu}\delta\phi_2~
\ee
with
 \be
 q\propto (-\eta)^{1/2-\nu}~{\rm and}  ~\frac{1}{2}-\nu=\frac{4-3(1+w_1)\alpha}{2(1+3w_1)}~.
 \ee
Defining $u=q\delta\phi_2$, it is easy to find that the Fourier transformation of $u$ satisfies the equation
\be
u_k''+(k^2-\frac{q''}{q})u_k=u_k''+(k^2-\frac{\nu^2-1/4}{\eta^2})u_k=0~.
\ee
With the selection of the Bunch-Davies vacuum at early time when $k|\eta|\gg 1$, the solution to the above equation is
\be
u_k=\sqrt{-\frac{\pi}{2}\eta} H_{\nu}^{(1)}(-k\eta)~,
\ee
where $H_{\nu}^{(1)}$ is the first kind Hankel function.  At late time when the perturbation mode is outside the Hubble radius, i.e.,
$k|\eta|\ll 1$, the mode function asymptotes $u_k\sim (-\eta)^{1/2} (-k\eta)^{-\nu}$. So that the power spectrum for $\delta\phi_2$ is
\be
\mathcal{P}_{\delta\phi_2}\sim k^3 |\frac{u_k}{q}|^2\sim k^{3-2\nu}~,
\ee
and does not depend on time. Hence the spectral index is
\be
n_s=4-2\nu=1+\frac{3(1+w_1)}{1+3w_1}(2-\alpha)~.
\ee
At later time this entropy perturbation will convert into the adiabatic perturbation which was constrained by the observations. The conversion
will not change the shape of the spectrum. The exact scale invariance corresponds to $\alpha=2$. Currently the observations show that the
spectrum has a tiny red tilt \cite{Planck1}, this requires the parameter $\alpha$ is larger than $2$ slightly. Different from the minimal
coupling case, here the scale invariance has nothing to do with the stability condition for the scaling solution, and also we do not need the
constraint $w_1\gg 1$.

Another issue concerns the non-Gaussianities of the primordial perturbations. Currently the Planck's result \cite{Planck} show that the
adiabatic perturbation has negligible non-Gaussianities. In the model considered here, the adiabatic perturbation is converted from
super-horizon entropy perturbation, so only the local type non-Gaussianity is significant \cite{Chen}. The local non-Gaussianity produced in
the models with non-standard kinetic terms was studied extensively in the literature. For example in the inflation scenario
the non-Gaussianity is estimated to be at the order of $f_{NL}\sim \mathcal{O}(1/c_{s2}^2)$ \cite{Chen}, here $f_{NL}$ is the estimator
of non-Gassianity, for the local shape it is defined in
\be
\zeta=\zeta_G+\frac{3}{5} f_{NL}[\zeta_G^2-<\zeta_G^2>]~,
\ee
where $\zeta$ is the curvature perturbation in the real space and $\zeta_G$ is the Gaussian curvature perturbation.
In our case even though the Ekpyrotic phase is quite different from inflation, the fluctuations of the spectator field live in an
effective de Sitter space, we will obtain the same estimation on the level of the non-Gaussianities.
As we mentioned before $c_{s2}^2=1$ if $K_2$ only depends on $X_2$, so the local non-Gaussianity is at most at the order of
$f_{NL}\sim \mathcal{O}(1)$. Furthermore, in our case the background of the spectator is frozen
$X_2=0$, the cubic action of $\delta\phi_2$ vanishes and the bispectrum of $\delta\phi_2$ are expected to be
negligibly small. Anyway during the Ekpyrotic phase no large non-Gaussianity is produced.
Another origin of the non-Gaussianity in the adiabatic perturbation is the nonlinear process which converts the entropy perturbation to the
adiabatic perturbation after the Ekpyrotic phase. This is model dependent. However, as long as the conversion is efficient, the resulted
non-Gassianity is expected to be at the order of $f_{NL}\sim O(1)$ as argued in  \cite{Fertig:2013kwa,Ijjas:2014fja}. So in these models the
primordial non-Gaussianities are controllable.

A simple example of the non-minimal coupling models can be constructed by a polynomial function for the kinetic term of the
spectator, i.e., $K_2=\sum_{n=1}^N c_n X_2^n$. Because $K_0=(dK_2/dX_2)|_{X_2=0}$ should be positive\footnote{A negative $K_0$
will lead to ghost instability. If $K_0=0$ there will be no dynamical equation for the perturbation $\delta\phi_2$.},
the coefficient of the first term should be positive, $c_1> 0$. Due to the frozen background $X_2=0$, the terms with higher powers will not
affect the background evolution and the linear perturbation. Other examples include $K_2\sim \exp(\lambda X_2)-1$ and
$K_2\sim \ln(1+\lambda X_2)$ with $\lambda>0$. One can show that in these simple cases we always obtain the sound speed $c_{s2}^2=1$.
More complicated models, for example $K_2=K_2(X_2, \Box\phi_2)$, deserve further explorations.

All the cases discussed above base on the scaling solution in which the equation of state of the universe is a constant 
during the Ekpyrotic phase. This is guaranteed by the function $1/\phi_1^2$ multiplying to $K_1(X_1)$ and $K_2(X_2)$ in the Lagrangian density. 
Similar to the generalization which has been done
for the models with canonical kinetic terms \cite{Ijjas:2014fja}, one can
perform a further generalization of the non-minimal coupling models studied in this paper to the case where the equation of state of the 
dominating field is not a constant. 
The models have the following action
\be\label{nonminimal1}
S=\int d^4x\sqrt{g}[\frac{R}{2}+P(\phi_1, X_1)+\Omega^2(\phi_1)K_2(X_2)]~,
\ee
where the two functions $P(\phi_1, X_1)$ and $\Omega(\phi_1)$ are arbitrary. 
As a simpler example, we may focus on the models with $P(\phi_1, X_1)=f(\phi_1)K_1(X_1)$ \footnote{One can see that 
this type of models includes the normal canonical model. 
Given the Lagrangian density of a scalar field $\psi$ with a canonical kinetic term
$\mathcal{L}=X_{\psi}-V(\psi)$, it
can be rewritten as $\mathcal{L}=f(\phi)(X_{\phi}-C)$ through the field redefinition $\phi=\int \sqrt{C/V(\psi)}d\psi$, 
here $f(\phi)=V(\psi)/C$ and $C$ is a constant.}. So the next purpose is to construct the functions $f(\phi_1)$ and $\Omega(\phi_1)$. 
The specific construction is model-dependent, but we can get the general idea following Ref. \cite{Ijjas:2014fja}. 
First we should have $\Omega'/\Omega+\mathcal{H}>0$ to freeze the spectator $\phi_2$, and the scale-invariance of the entropy 
perturbation requires $a\Omega\propto 1/(-\eta)$. 
As in Ref. \cite{Ijjas:2014fja}, given the form of time dependence of the fast-roll parameter 
$\epsilon(\eta)=(3/2)(1+w_1)$ we can obtain the functional form of the conformal Hubble parameter 
$\mathcal{H}(\eta)$ through the 
relation $\epsilon =1-\mathcal{H}'/\mathcal{H}^2$. Then the scale factor $a(\eta)$ is obtained by
\be
a(\eta)=a(\eta_{\rm end})\exp (\int^{\eta_{\rm end}}_{\eta} \mathcal{H}d\eta)~,
\ee
where $\eta_{\rm end}$ is the conformal time when the Ekpyrotic phase ends, it is convenient to set $\eta_{\rm end}=-1$. After that
one gets the time dependence of $\Omega$ through the requirement $\Omega(\eta)\propto 1/[-a(\eta)\eta]$. 
Furthermore we can use the relation 
\be
w_1=\frac{2\epsilon}{3}-1=\frac{K_1}{-K_1+2X_1 (dK_1/dX_1)}~,
\ee
and the Friedmann equation
\be
\mathcal{H}^2=\frac{a^2}{3} f(\phi_1)(-K_1+2X_1 \frac{dK_1}{dX_1})~,
\ee
to obtain the forms of $f(\eta)$ and the field $\phi_1(\eta)$ as functions of time. This process is model-dependent,
it depends on the specific form of the function $K_1(X_1)$. Once this is known we can invert $\phi_1(\eta)$ to get $\eta(\phi_1)$ and then 
substitute it into the obtained $f(\eta)$ and $\Omega(\eta)$ to get the expressions $f(\phi_1)$ and $\Omega(\phi_1)$. 
With little modification, this construction works equally well for the case where the spectral index of the 
entropy perturbation differs from the exact scale-invariance by a constant tilt. After the constructions of $f(\phi_1)$ and $\Omega(\phi_1)$, 
one must check whether the solution $(\phi_1, \phi_2={\rm const.})$ is stable. But this is again model-dependent.

\section{Conclusions}

The Ekpyrotic universe as an alternative to inflation should provide not only solutions to the problems of the big bang cosmology but also a
mechanism to generate primordial perturbation for structure formation. Currently the best way to generate the primordial perturbations
consistent with observations is the entropic mechanism. In cases of canonical kinetic terms, the entropic models with minimal couplings
encountered the difficulties of instability \cite{koyama} and possible large non-Gaussianities. These difficulties can be overcome in the
models with non-minimal couplings \cite{Li:2013hga,Ijjas:2014fja,qiu1}. In this paper we extended these studies to the cases of non-standard
kinetic terms. The field models with non-standard kinetic terms have extensive applications in cosmology and are expected from fundamental
theories. We first proved that in the framework of non-standard kinetic terms, the minimal coupling entropic mechanisms are still unstable,
the condition of stability conflicts with the requirement for scale invariant perturbations.  Then we discussed the non-minimal coupling
entropic mechanisms. We showed that for some simple cases the non-minimal coupling models can be stable and produce nearly scale invariant
and Gaussian primordial perturbations.

\section{Acknowledgement}

The author thanks Taotao Qiu for useful discussions. This work is supported by NSFC under Grant No. 11422543 and by Program for New Century
Excellent Talents in University.

{}

\end{document}